\documentstyle[12pt]{article}

\makeatletter
\@addtoreset{equation}{section}
\makeatother

\let\a=\alpha \let\b=\beta   \let\e=\epsilon
    
\let\l=\lambda     \let\r=\rho
\let\s=\sigma

     \let\la=\label
   
 \def\nn{\nonumber} \def\bd{\begin{document}}
\def\ed{\end{document}} \def\ds{\documentstyle} \let\fr=\frac
\let\bl=\bigl \let\br=\bigr \let\Br=\Bigr \let\Bl=\Bigl \let\bm=\bibitem
\let\na=\nabla \let\pa=\partial \let\ov=\overline
\newcommand{\be}{\begin{equation}} \newcommand{\ee}{\end{equation}}
\def\ba{\begin{array}} \def\ea{\end{array}}
\newcommand{\bea}{\begin{eqnarray}} \newcommand{\eea}{\end{eqnarray}}
\newcommand{\ra}{\rightarrow} \newcommand{\oh}{\frac{1}{2}}
\newcommand{\lra}{\longrightarrow} \newcommand{\Lra}{\Leftrightarrow}
\newcommand{\ap}{\alpha^\prime} \newcommand{\bp}{\tilde \beta^\prime}
\newcommand{\tr}{{\rm tr} } \newcommand{\sg}{\sqrt {-g}}
\newcommand{\rn}{r_0} \newcommand{\rpl}{r_p} \newcommand{\hoch}[1]{$\,
^{#1}$} \newcommand{\srt}{\sqrt{3}}
\newcommand{\oosrt}{\frac{1}{\sqrt{3}}} \newcommand{\rmi}{r_-}
\newcommand{\Dp}{\Delta_{+}} \newcommand{\Dm}{\Delta_{-}}
\newcommand{\Tr}{{\rm Tr} } \newcommand{\NP}{Nucl. Phys. }
\newcommand{\lag}{{\cal L}} \newcommand{\tamphys}{\it Center for
Theoretical Physics\\ Physics Department \\ Texas A \& M University \\
College Station, Texas 77843} \newcommand{\auth}{M.~J.
Duff\hoch{\dagger} and J. Rahmfeld}

\newcommand{\dt}{\tilde{d}}
\newcommand{\ratio}{\frac{d\tilde{d}}{d+\dt}}
\newcommand{\vaa}{\vec{a}_{\a}}
\newcommand{\vp}{\vec{\phi}}

\begin{document}
\hfill{}
\hfill{}

\hfill{CTP-TAMU-17/96} 

\hfill{hep-th/9605085}

\vspace{24pt}

\begin{center}
{ \large {\bf Bound States of Black Holes and Other $P$-branes}}

\vspace{48pt}

\auth

\vspace{10pt}

{\tamphys}

\vspace{72pt}

\underline{ABSTRACT} 
\end{center} 
In the process of identifying heterotic and Type $II$ BPS string
states with extremal dilaton black holes, it has been suggested that
solutions with scalar/Maxwell parameters $a=\sqrt{3}$, $1$,
$1/\sqrt{3}$ and $0$ correspond to $1-$, $2-$, $3-$ and $4$-particle
bound states at threshold. (For example, the Reissner-Nordstrom black
hole is just a superposition of four Kaluza-Klein black holes). Here
we show that not only the masses, electric charges and magnetic
charges but also the spins and supermultiplet structures of the string
states are consistent with this interpretation. Their superspin $L$
corresponds to the Kerr-type angular momentum and hence only the $L=0$
elementary BPS states are black holes. Moreover,
these results generalize to super $p$-branes in $D$-dimensions.  By
constructing multi-centered $p$-brane solitons, the new super $p$-branes found
recently with various values of $a^2=\Delta-2(p+1)(D-p-3)/(D-2)$ are seen to
be bound states of the fundamental ones with $\Delta=4$.

{\vfill
\leftline{}
\vskip 10pt
\footnoterule
{\footnotesize
 \hoch{\dagger} Research supported in part by NSF Grant PHY-9411543}

\newpage
\section{Introduction}
\la{intro}

In a previous paper \cite{Rahmfeld1}, we argued that the spectrum of
elementary BPS ($N_R=1/2$) states of compactified heterotic strings could be
identified with extremal electrically charged black holes. Further evidence
for this interpretation was supplied in
\cite{Sen2,Peet,Khurimyers,Callan,Shiraishi}. In particular, the $N_L=1$ states
and the $N_L>1$ states (with vanishing left-moving internal momentum) admit a
single scalar/Maxwell interpretation with parameters $a=\sqrt{3}$ or $a=1$
respectively.  In other words, by choosing appropriate combinations of dilaton
and moduli fields to be the scalar field $\phi$ and appropriate combinations
of the field strengths to be the Maxwell field $F$, the field equations can be
consistently truncated to a form given by the Lagrangian  \be {\cal L}=
\frac{1}{32\pi}\sqrt{-g}\left [R-\frac{1}{2}(\partial \phi)^2-\frac{1}{4}e^{-a
\phi}F^2 +...\right] \la{actiona}  \ee 
for these two values of $a$, these combinations being just those
corresponding to the quantum numbers of the string states\footnote{A {\it
consistent} truncation is defined to be one for which all solutions of
the truncated theory are solutions of the original theory. The dots in
(\ref{actiona}) refer to terms involving a pseudoscalar combination of axion and
moduli fields which are in general required for consistency but which do not
contribute to non-rotating black hole solutions.  We are grateful to R. Myers
for pointing out their omission in the original version of this paper.}. In
the case of zero angular momentum, the ADM mass $M{}_{black}$ of the extremal
black hole solutions of (\ref{actiona}) is given by  \be M_{black}^2=Q^2/4(1+a^2)
\la{2} \ee  where $Q=\int \tilde F/8\pi=\int e^{-a\phi}{}{*F}/8\pi$ is the
electric charge, where  $*$ denotes the Hodge dual and where, for simplicity, we
have set the asymptotic value of $\phi$ to zero. The $a=1$ case yields the
supersymmetric dilaton black hole \cite{Gibbons}. The $a=\sqrt{3}$ case
corresponds to the Kaluza-Klein black hole and the ``winding" black hole
\cite{Khurinew} which are related to each other by $T$-duality. The Kaluza-Klein
solution has been known for some time \cite{Gibbons} but only recently
recognized \cite{Khurinew} as a heterotic string solution. We also argued that
the corresponding solitonic magnetically charged and dyonic spectrum, predicted
by S-duality \cite{Schwarz1}, is also described by extremal black holes. Indeed,
we were first motivated to make the black hole conjecture for the elementary
states by first noting that the ``winding'' magnetic monopoles are extremal
black holes \cite{Khurinew} and then noting that string/string duality
interchanges the roles of $S$-duality and $T$-duality and therefore that the
solitonic monopoles play the same role for the dual string as the elementary
winding states play for the fundamental string \footnote{Of course this involves
extending the classical notion of a black hole from the weak coupling to the
strong coupling regime.  We will therefore take the liberty of  describing a
state by the words {\it black hole} if there exists at least one string picture
in which its mass exceeds the Planck mass for weak coupling.}.  By allowing $F$
to describe combinations of field strengths and their duals, other dyons,
preserving fewer supersymmetries and therefore not predicted by S-duality alone,
can also be assigned $a$ values. One finds $a=1/\sqrt{3}$ and $a=0$. The $a=0$
case yields the Reissner-Nordstrom solution\footnote{The Reissner-Nordstrom
black hole is not a solution of dimensionally reduced pure gravity but has long
been known to be a solution of $M$-theory \cite{Duffpope,Pope}.} which,
notwithstanding contrary claims in the literature, does solve the low-energy
string equations \cite{Khurinew,Rahmfeld1}. The $a=1/\sqrt{3}$ black hole
\cite{Horowitz2} was identified as a dyonic solution in \cite{Rahmfeld3}. These
four values of $a$ yield solutions which are special cases of the most general
solution subsequently found in \cite{Cveticyoum}, and shown to be exact to all
orders in $\alpha'$ in \cite{Cvetictseytlin}. In the $N=2$ theory, black holes
with $a=\sqrt 3,1,1/\sqrt 3,0$ all preserve $1$ supersymmetry and therefore
belong to fundamental supermultiplets with maximum spins $1/2$ \cite{Rahmfeld3}.
In the $N=4$ theory, black holes with $a=\sqrt{3},1,1/\sqrt{3},0$ preserve
$2,2,1,1$ supersymmetries, respectively, and therefore belong to fundamental
supermultiplets with maximum spins $1,1,3/2,3/2$
\cite{Kalloshpeet,Rahmfeld1,Rahmfeld3}.  This same black hole interpretation
could be extended to the spectrum of BPS states of toroidally compactified Type
$II$ strings \cite{Hull} where the supersymmetric black holes admitting a single
scalar/Maxwell interpretation correspond once again to the same four values of
$a=\sqrt{3},1,1/\sqrt{3},0$. They preserve $4,2,1,1$ of the $N=8$
supersymmetries, respectively \cite{Popelu1,Khuriortin}, and therefore belong to
fundamental supermultiplets with maximum spins $2,3,7/2,7/2$. 

On the basis of their mass and charge assignment, it was further
suggested \cite{Rahmfeld1,Rahmfeld3} that we interpret these four values
of $a$ as $1-, 2-, 3-$ and $4$-particle bound states with zero binding
energy. This is reviewed in section (\ref{bound}).  For example, the
Reissner-Nordstrom ($a=0$) black hole equals four Kaluza-Klein
($a=\sqrt{3}$) black holes! This zero-binding-energy bound-state
conjecture can, in fact, be verified in the classical black hole picture
by finding explicit $4$-centered black hole solutions which coincide
with the $a=\sqrt{3},1,1/\sqrt{3},0$ solutions as we bring $1,2,3,4$
centers together and take the remaining $3,2,1,0$ centers out to infinity
\cite{Rahmfeld4}. Such a construction is possible because of the
appearance of four independent harmonic functions
\cite{Cvetictseytlin}. Moreover, this provides a novel realization of the {\it
no-force} condition in that the charge carried by each black hole corresponds
to a different $U(1)$. Thus the gravitational attraction cannot be cancelled
by an electromagnetic repulsion but rather by a subtle repulsion due to scalar
exchange. This phenomenon was also observed in \cite{Kalloshcancel}. In
section (\ref{supermultiplets}) we shall provide further evidence for the
bound state interpretation in the quantum string state picture by showing that
not only the masses, electric charges and magnetic charges but also the spins,
and supermultiplet structures are consistent with this interpretation of the
$a=0,1/\sqrt 3,1$ string states being merely bound states of the fundamental
$a=\sqrt{3}$ states. This is entirely consistent with the claim of
\cite{Holzhey}, using completely different reasoning, that only $a>1$ dilaton
black holes can be interpreted as {\it elementary} particles. Can this
interpretation also apply to the recently discussed {\it massless} black holes
\cite{Strominger2,Cveticyoum2,Behrndt,Kalloshcancel,Chancvetic}? Classically, the
answer is yes in that there exist 2-centered solutions which coincide
with the massless black hole as we bring the two centers together. In
this case, however, it is necessary to assume that one of the constituents
has a negative mass and it therefore seems unlikely that this bound state
interpretation can survive quantum-mechanically.  

For the purely electric elementary string states, where it makes sense to
assign an oscillator number, the massive BPS states in the
heterotic theory are given by the $(N_R=1/2,N_L\geq1)$ states
\cite{Rahmfeld1} and the massless  BPS states belong to the $(N_R=1/2,N_L=0)$
sector \cite{Behrndt}. Curiously, however, it is possible to extend the black
hole bound state interpretation to non-BPS states, for example the
non-supersymmetric $a=1$ dilaton black hole of \cite{Garfinkle} corresponds to
$(N_R=3/2,N_L=1)$.

There is now a consensus that all of string theory and its duality
properties follow from an underlying eleven-dimensional theory
\cite{Howe,Townsend,Witten,Duffliuminasian,Schwarz2,Horava}, now known as
{\it $M$-theory}. In section (\ref{multi}) we turn to the black branes of
$M$-theory \cite{Dufflupope2,Klebanov1,Mukherji2,Muto}, where (\ref{actiona})
is now in an arbitrary spacetime dimension $D\leq11$ and the $F$ is now a
$(p+2)$-form.  The parameter $a$ can be conveniently re-expressed as
\cite{Popestainless}  \be a^2 = \Delta - \frac{2 (p+1)(D-p-3) }{D-2} , \ee
since $\Delta$ is a parameter that is preserved under dimensional reduction
\cite{Popestainless}.  Originally, attention was focussed on the $\Delta=4$
solutions \cite{Strominger1,Dufflu,Khuristring} but various new supersymmetric
solitons with $\Delta\neq 4$ have recently been studied
\cite{Popestainless,Popelu1,Popelu2}.  These authors proposed to classify
$p$-branes into ``rusty'' or ``stainless'' according as they can or cannot be
``oxidized'' to isotropic brane solutions of a higher-dimensional
supergravity. (Oxidation is the opposite of double dimensional reduction
\cite{Howe,fifteen}.) Examples of new stainless solutions included a $\Delta=2$
$5$-brane in $D=9$ and  $\Delta=2,4/3$ strings in $D=5$ \footnote{The 
$\Delta=4$ $6$-brane in $D=9$ can, in fact, be oxidized to a Type $IIB$
$7$-brane in $D=10$.}. The authors of \cite{Popestainless} then raised the
question of whether these new $\Delta \neq 4$ branes deserve to be treated as
fundamental in their own right. In this section we shall generalize the
treatment of extremal black holes \cite{Rahmfeld4} and confirm on the level of
classical solutions that these new $\Delta=4/n$ $p$-branes can also be
regarded as bound states with zero binding energy of $n$ fundamental
$\Delta=4$ branes.   We find new $1\leq n \leq m$-centered $p$-brane solutions
which reproduce the $\Delta=4/n$  solutions of \cite{Popelu2} as we allow $n$
of the centers to coincide and take the remaining $(m-n)$ out to infinity. In
particular, the $\Delta=2$ fivebrane is a bound state of two $\Delta=4$
fivebranes and the  $\Delta=2$ and $\Delta=4/3$ strings are respectively bound
states of two and three fundamental $\Delta=4$ strings.   

In section (\ref{entropy}), we also discuss the temperature and entropy
of the extreme $a=\sqrt{3}$, $1$, $1/\sqrt{3}$, $0$ black holes. Here a good
deal more needs to be understood since only for the $a=0$ solution is the
dilaton vanishing and so the classical entropy prediction for the other
three cases (namely zero) is unreliable.  Similar remarks apply to the 
macroscopic entropy and temperature of the other $p$-branes. 

\section{The bound state conjecture}
\la{bound}

Let us begin by recalling the bound state conjecture in the context of
the four-dimensional heterotic string obtained by toroidal
compactification. At a generic point in the moduli space of vacuum
configurations the unbroken gauge symmetry is $U(1)^{28}$ and the low
energy effective field theory is described by $N=4$ supergravity
coupled to $22$ abelian vector multiplets. Using the canonical metric,
the bosonic part of the Lagrangian is given by \cite{Sen}
\begin{equation}
{\cal{L}}={1\over32\pi}\sqrt{-g}\left[R-\frac{1}{2}(\partial\eta)^2
-\frac{1}{12}e^{-2\eta}H^2+{1\over8}\Tr(\partial  {\cal M}
L\partial {\cal M} L)-{1\over4}e^{-\eta}F{}^T
( L {\cal M} L)F\right],
\la{full}
 \end{equation}
where $L$ is the metric of $O(6,22)$. ${\cal M}={\cal M}^T\in O(6,22)/O(6)\times
O(22)$ parametrizes the scalars in the sigma model, the 28
$F_{\mu\nu}$'s are the $U(1)$ fields strengths, $\eta$ is the four
dimensional dilaton and $H$ the $3$-form field strength with the usual
Chern-Simons terms. The string theory has a perturbative $O(6,22;Z)$ $T$-duality
that transforms Kaluza-Klein states into winding states, and a non-perturbative
$SL(2,Z)$ $S$-duality that transforms electric states into magnetic states. 
This is reflected in the $O(6,22;R)$ invariance of the Lagrangian (\ref{full})
and the $SL(2,R)$ invariance of its equations of motion.  

Let us work at a special point in the moduli space and set the
asymptotic value of ${\cal M}$ to $I$, and the asymptotic dilaton field
to zero. We shall return to the general case later.
Let us define $\vec{Q}_{R,L}$ and $\vec{P}_{R,L}$ by
\bea
\vec{Q}_{R,L}&=&\frac{1}{2}(I\pm L)\vec{Q} \nn \\
\vec{P}_{R,L}&=&\frac{1}{2}(I\pm L)\vec{P}\ ,  
\eea
where $\vec{Q}$ and $\vec{P}$ are the $28$-dimensional electric and
magnetic charge vectors. Denoting by $N_L$ and $N_R$ the number of
left and right oscillators respectively, the mass of an elementary
(purely electric) string state is given by 
\be
M_{string}^2=\frac{1}{8}(\vec{Q}_R^2+2 N_R-1)=\frac{1}{8}\ 
  (\vec{Q}_L^2+2 N_L-2)
\la{stringmass} 
\ee
See, for example, \cite{Sen}.  In this $N=4$ theory, states (whether
elementary or solitonic) fall into $3$ categories according as they are
annihilated by $q=2,1,0$ supersymmetries, in which case their masses are
given by \cite{Rahmfeld3} 
\bea
M_{central}&=&Z_1=Z_2~~~~~~~q=2\nn\\
&=&Z_1>Z_2~~~~~~~q=1\nn\\
&>&Z_1 \geq Z_2~~~~~~~q=0
\eea
where $Z_1$ and $Z_2$ are the moduli of the 
central charges in the supersymmetry
algebra given by \cite{Cveticyoum,Rahmfeld3} 
\begin{equation}
Z_{1,2}^2=\frac{1}{8}\left[\vec{Q}_R^2+\vec{P}_R^2\pm
2\left(\vec{Q}_R^2\vec{P}_R^2-(\vec{Q}_R\vec{P}_R)^2\right)^{\frac{1}{2}}\right]\
. \la{central} 
\end{equation}
It follows by comparing $M_{string}$ of (\ref{stringmass}) and
$M_{central}$ of (\ref{central}) that the elementary string states, being
purely electric, are either BPS states with $N_R=1/2$ or else
non-supersymmetric states with $N_R >1/2$.  

The $N_R=1/2$ states correspond to that subset of the full spectrum that
is annihilated by half of the supersymmetry generators ($q=2$), belongs
to short representations of the $N=4$ supersymmetry algebra and
saturates the Bogomol'nyi bound $M_{central}=Z_1=Z_2$. The basic
superspin $L=0$ multiplet is the $16$ dimensional ($J_{max}=1$)
multiplet $(1,4,5)$. This is the only multiplet appearing for $N_L=0$.
However, higher values of superspin $L$ may appear for higher $N_L$.
Since the left moving oscillators have spins $0$ if they lie in the
$22$ compact dimensions or $1$ if they lie in the spacetime dimensions,
the superspin obeys the bound $L\leq N_L$. In particular for $N_L=1$, we
have in addition to the above $L=0$ multiplet the $48$ dimensional   
($J_{max}=2$) multiplet $(1,4,6,4,1)$. For $N_L>1$ we have $J_{max}
=L+1$.

The $N_R >1/2$ states, belonging to the long 
representations, are annihilated by no supersymmetries ($q=0$) and
satisfy $M_{central}>Z_1 \geq Z_2$. The $L=0$ multiplet is $256$
dimensional with ($J_{max}=2$).  No elementary states belong to
the intermediate  representation,
which are annihilated by one supersymmetry ($q=1$) and satisfy
$M_{central}=Z_1>Z_2$. These $L=0$ multiplets are $64$ dimensional
with ($J_{max}= 3/2$).  The situation changes when we allow for
solitonic states of the string theory which carry magnetic charge. Then
we can have all three categories of supermultiplet
\cite{Rahmfeld1,Rahmfeld3}, and indeed we predicted the existence of
new dyonic ($J_{max}=3/2$) multiplets in the string spectrum, over and
above the ($J_{max}=1$) dyonic states related by $S$-duality to the
elementary states and predicted by Schwarz and Sen \cite{Schwarz1}.

In \cite{Rahmfeld1} we considered these elementary electrically charged
massive $N_R=1/2$ states, and showed that the spin zero, superspin
zero, states correspond to extreme limits of non-rotating black hole
solutions which preserve $1/2$ of the spacetime supersymmetries. By
supersymmetry, the black hole interpretation then applies to all
members of the $N=4$ supermultiplet \cite{Gibbonsperry,Aichelburg},
which has $J_{max}=1$. For a subset of states the low-energy string
action can be truncated to (\ref{actiona}). The scalar-Maxwell
parameter is given by $a=\sqrt 3$ for $N_L=1$ and $a=1$ for $N_L>1$
(and vanishing left-moving internal momenta). The other superspin
zero states with $N_L>1$ are extremal non-rotating black holes too, but
are not described by a single scalar truncation of the type
(\ref{actiona}). We also made a similar identification for the dyonic
states \cite{Rahmfeld3}. To see this, let us recall that general black
hole solutions are determined by the $56$ components of the electric and
magnetic charge vectors $\vec{Q}$ and $\vec{P}$. We can simplify the
problem by applying an $O(6)\times O(22)$ T-duality transformation
which eliminates all but four electric and four magnetic components.
This corresponds to a truncation involving just four field strengths:
$F_1, \ F_2, \ F_3$ and $F_4$, and just three complex scalars:
$S=a+ie^{-\eta}$, $T=b+ie^{-\sigma}$, $U=c+ie^{-\rho}$.  \bea 
{\cal L}&={1\over 32\pi}\sqrt{-g}\biggl \{
&\hspace{-0.4truecm}R-\frac{1}{2}\left[(\partial\eta)^2
+(\partial\s)^2+(\partial\rho)^2\right]-\nn\\ & &
-\frac{1}{2}\left[e^{2\s}(\partial b)^2 +
e^{2\r}(\partial c)^2\right]-\frac{1}{12}e^{-2\eta}H^2- \nn \\ &&
\hspace*{-0.5truecm} -\frac{e^{-\eta+\s+\r}}{4} \biggl[
|T|^2 |U|^2 F_1^2 +|T|^2F_2^2+ F_3^2+ |U|^2 F_4^2
+2bF_2F_3+2cF_3F_4 +\nn\\ &&\hspace{20mm}-2c
|T|^2F_1F_2 -2b|U|^2 F_1F_4 -cb (F_1F_3-F_2F_4)\biggr]\biggr\}, \la{action1} 
\eea 
It is noteworthy that
this procedure keeps only the degrees of freedom which arise from the
toroidal compactification from six to four dimensions: $N=2$
supergravity coupled to three vector multiplets.  $F_1,F_2$ are the
Kaluza-Klein fields, $F_3,F_4$ are the winding fields: $S$ is the
axion/dilaton, $T$ the Kahler-form and $U$ the complex structure. The axion
field is obtained by dualizing $H$, which satisfies the standard Bianchi
identity. The Lagrangian (\ref{action1}) was thoroughly analyzed in
\cite{Rahmfeld3} where, in particular the triality of the three duality
groups, $SL(2;Z)_S$, $SL(2;Z)_T$, $SL(2;Z)_U$, was emphasized: (\ref{action1})
is obviously invariant under $T\leftrightarrow U, \ F_2\leftrightarrow F_4$
exchange, but the equations of motion are also invariant under
$S\leftrightarrow T,\ S\leftrightarrow U $ exchange (accompanied by an
appropriate electric/magnetic transformation of the field strengths), which
lead to the interpretation of a triality of the $S$, $T$ and $U$ strings.
To simplify further, we shall here consider solutions with vanishing
pseudoscalars so that the reduced Lagrangian is 
\bea 
{\cal L}&={1\over 32\pi}\sqrt{-g}\biggl \{
&\hspace{-0.4truecm}R-\frac{1}{2}\left[(\partial\eta)^2
+(\partial\s)^2+(\partial\rho)^2\right]-\nn\\ & & \hspace*{-0.5truecm}
-\frac{e^{-\eta}}{4}\left[e^{-\s-\r}F_1^2+e^{-\s+\r}F_2^2
+e^{\s+\r}F_3^2 +e^{\s-\r}F_4^2\right] \biggr\}, \la{action}
\eea 
where
one has to keep in mind the constraints imposed by the requirement of
vanishing pseudoscalars.

The extremal black holes we have in mind to illustrate the point,
namely those that allow for a description by an action of type
(\ref{actiona}) with $a=\sqrt{3},1,1/\sqrt{3},0$, can be obtained
from (\ref{action}) by making various truncations:
\bea
a=\sqrt{3}: & F_{1}\neq0,  &F_2=F_3=F_4=0,~~Q^2=1 \nn \\
a=1: & F_{1}=F_3\neq0, & F_2=F_4=0,~~~~~~~~~Q^2=2 \nn \\
a=1/\sqrt{3}: & F_{1}=F_3={\tilde F}_2\neq0, & F_4=0,~~~~~~~~~~~~~~~~Q^2=3
\nn \\ a=0: & F_{1}=F_3={\tilde F}_3 ={\tilde F}_4
\neq0, &~~~~~~~~~~~~~~~~~~~~~~~~~~Q^2=4 \eea 
Table 1 summarizes the
charge and mass quantum numbers for
those and a few more 
black holes and string states in the heterotic string theory. (If so desired,
one may then use  $S-T-U$ triality to describe them in the dual Type
$II$ string pictures \cite{Rahmfeld3}.) Based on this Table it can be easily
verified \cite{Rahmfeld1,Rahmfeld3} that the mass and charge quantum numbers
of the $a=0,1/\sqrt{3},1,\srt$ black holes admit the interpretation  of
$4,3,2,1$ particle bound state with zero binding energy. For the purposes of
illustration we have chosen the special cases where all non-zero charges are
equal to unity but it is easily generalized to the case of different charges
$Q_1$, $P_2$, $Q_3$ and $P_4$ \cite{Cveticyoum} where the interpretation is
that of a $(Q_1+P_2+Q_3+P_4)$-particle bound state with zero binding energy. 

Note, by the way, that even the extremal, but {\it non-supersymmetric},
$a=1$ black hole with electric charges $(1,0,-1,0)$ fits into the string
spectrum with the assignments $N_L=1,N_R=3/2$. (Since the masses of
non-supersymmetric states are not protected from quantum corrections by
any non-renormalization theorems we do usually not expect
(\ref{stringmass}) to give the correct answer for arbitrary black
holes.)  This solution is related by T-duality \cite{Rahmfeld1} to the
non-supersymmetric black hole of \cite{Garfinkle} which has just one
non-vanishing charge corresponding to one of the $16$ $U(1)$s in the
Yang-Mills sector. It is frequently claimed that extremal black holes and
supersymmetric black holes are synonymous, but while
$M_{string}^2={\vec{Q}}_L^2/8$ and  $M_{string}^2={\vec{Q}}_R^2/8$ are
both extremal, only $M_{string}^2={\vec{Q}}_R^2/8$ is supersymmetric,
owing to the left/right asymmetry of the heterotic string.  (Note, 
however, that this solution preserves one quarter of the supersymmetries
when embedded into maximal $N=8$ supergravity coming from the Type $II$
string). {\footnotesize \begin{table} \begin{center}
\begin{tabular}{|c|c|c||c|c|c|c||c|c|} \hline
    \multicolumn{3}{|c||} {\em Quantum Numbers} & \multicolumn{4}{c||} 
   {\em String  States} & \multicolumn{2}{c|} {\em Black Holes }\\ 
  \hline \hline
     {$(Q_1,Q_2,Q_3,Q_4)$} & {$
 (P_1,P_2,P_3,P_4)$} &q &$N_L$ &$N_R$ &$M_{string}^2$ 
     & $M_{central}^2$ & a &  $M_{black}^2$   \\ \hline 
\hline
   & & & & & & & & \\
   (1,0,0,0)  & 
   (0,0,0,0) & 2 &1 & $\frac{1}{2}$&
    $\frac{1}{16}$ & $\frac{1}{16}$ &  $\sqrt{3}$ & $\frac{1}{16}$\\
   (1,0,1,0)  & (0,0,0,0) &2 &2 & $\frac{1}{2}$&
    $\frac{1}{4}$ & $\frac{1}{4}$ & 1 & $\frac{1}{4}$ \\
   (2,1,1,1)  & (0,0,0,0) &2  &4 & $\frac{1}{2}$&
    $\frac{13}{16}$ &  $\frac{13}{16}$ &x& $\frac{13}{16}$ \\
 (1,0,-1,0)  & (0,0,0,0) & 2& 0 & $\frac{1}{2}$&
     $0$ & 0 & x &$0$ \\ 
 (1,0,-1,0)  & (0,0,0,0) & 0& 1 & $\frac{3}{2}$&
     $\frac{1}{4}$ & 0 & 1 &$\frac{1}{4}$ \\ 
   & & & & & & & & \\
     (1,0,0,0)  & (0,1,0,0) & 1&x & x&
     x & $\frac{1}{4}$ & 1& $\frac{1}{4}$ \\
   (1,0,1,0)& (0,1,0,0)& 1& x& x& x 
    & $\frac{9}{16}$ &  $\frac{1}{\sqrt{3}}$ & 
    $\frac{9}{16}$ \\
    (1,0,1,0)  & (0,1,0,1) &1& x & x & x & $1$ & 0 & 1 \\  
 (1,0,-1,0)  & (0,1,0,-1) & 0& x & x &
     x &  0 & 0 &
     1 \\\hline  
\end{tabular}
\end{center}
\normalsize
\begin{center}Table 1: Masses and charges of black holes and
string states. \end{center}
\end{table}
\normalsize}

As discussed in \cite{Rahmfeld1}, although the superpartners of the
non-rotating black holes are themselves black holes, they are {\it not}
rotating black holes in the sense of Kerr. On the contrary, as
explained in \cite{Aichelburg}, it is their fermionic hair that carries
the angular momentum in contrast to conventional rotating black holes
where the angular momentum is bosonic. Rather this bosonic angular
momentum is supplied by the left moving oscillators, which leads us to
identify the Kerr-type angular momentum with the superspin $L$.
However, as also discussed in \cite{Rahmfeld1}, these $N_R=1/2$ string
states cannot then be rotating {\it black holes} since these
mass=charge solutions have event horizons only for vanishing Kerr angular
momentum.

\section{Black
hole supermultiplets}  \la{supermultiplets}

\begin{table}
\begin{center}
\begin{tabular}{|c||ccccc|}\hline
 & &$a=0$&$a=1/\sqrt{3}$&$a=1$&$a=\sqrt{3}$\\ \hline\hline
$N=2$&&1&1&1&1\\
$N=4$&&1&1&2&2\\
$N=8$&&1&1&2&4\\ \hline
\end{tabular}
\end{center}
\normalsize
\begin{center}Table 2:
Number of preserved supersymmetries, $q$, for black holes with
parameters $a=0,1\sqrt{3},1,\sqrt{3}$ in $N=2,4,8$ theories.
\end{center}
\end{table}
\normalsize
Since the basic quantities like mass and charge are not going to change if we
move to the $N=8$ or $N=2$ theories, we may as well extend the conjecture to
these theories also.  The number of preserved supersymmetries is given in
Table 2.  In this section we check that the supermultiplet structure of the black
holes is consistent with this bound state interpretation. The relevant group
theory may be found in \cite{Ferrarasavoy}. Let us recall that an $N$-extended
supersymmetry algebra admits $N/2$ central charges $Z_1,...,Z_{N/2}$. States
fall into $1+N/2$ categories according as they are annihilated by $ N/2 \geq q
\geq 0$ supersymmetry generators. $q$ also counts the number of $Z$s that obey
the bound $M_{central}=Z_{max}$. We are primarily interested in multiplets with
states with spin $J=0$ and superspin $L=0$, which we identify with the
non-rotating black hole solutions. The rest of the $L=0$ supermultiplet may then
be filled out using the fermionic zero modes \cite{Aichelburg}. In the
fundamental multiplets the spin will run from $J=0$ up to $J= (N-q)/2$ whereas
for multiplets with non-zero superspin $L$, the spin will run from $J=0$ to
$J=L+(N-q)/2$ for $L<(N-q)/2$ and  from $J=L-(N-q)/2$ to $J=L+(N-q)/2$ for $L
\geq (N-q)/2$. The multiplets are constructed in the spirit of
\cite{Ferrarasavoy} by combining massless supermultiplets and then employing the
Higgs mechanism to obtain the massive multiplet. For each number $q$ of
preserved supersymmetries we give the results up to $J_{max}=4$ (implying
superspins $L=0,\frac{1}{2},1,\frac{3}{2}$ and $L=2$ for the case of four
preserved supersymmetries and so on). The results are shown in Tables 3, 4 and 5
for the $N=8, \ N=4$ and $N=2$ theories, respectively.

{\footnotesize
\begin{table}
\begin{center}
\begin{tabular}{|r |r| r r r r r |}
\hline $q$&$J$&$L=0$&$L=\frac{
1}{2}$& $L=1$&$L=\frac{3}{2}$&
$L=2$\\
\hline\hline
4 & 4                 &     &   &      &    &1\\ 
 & $\frac{7}{2}$  &    &    &      &1   &8\\ 
 &3                    &     &   & 1   &8   &28\\ 
 &  $\frac{5}{2}$ &    & 1 &8    & 28&56\\ 
 & 2                   & 1  & 8 & 28 & 56&70\\ 
 &  $\frac{3}{2}$ & 8  & 28& 56& 70&56\\ 
 & 1                   & 27 & 56& 70& 56&28\\ 
 & $\frac{1}{2}$  &48  & 69& 56& 28&8\\ 
 &0                    & 42 & 48& 27&8   &1\\ \hline
3 & 4                      &    &      &    &1&\\ 
 & $\frac{7}{2}$      &     &      &1   &10&\\ 
 &3                         &    & 1   &10   &45&\\ 
 &  $\frac{5}{2}$     & 1   & 10    & 45&120&\\ 
 & 2                       & 10 &  45 & 120&210&\\ 
 &  $\frac{3}{2}$     & 44 & 120& 210&252&\\ 
 & 1                       & 110& 209& 252&210&\\ 
 & $\frac{1}{2}$  &     165& 242& 209&120&\\ 
 &0                        & 132& 165&110   &44&\\ \hline
2 & 4                     &&   &1&&\\ 
 & $\frac{7}{2}$      &  &   1   &12&&\\ 
 &3                       & 1   &12  &66&&\\ 
 &  $\frac{5}{2}$   & 12    & 66&220&&\\ 
 & 2                     &  65 & 220&495&&\\ 
 &  $\frac{3}{2}$ &     208& 494&792&&\\ 
 & 1                      & 429& 780&923&&\\ 
 & $\frac{1}{2}$    & 527& 858&780&&\\ 
 &0                     & 429 & 572   &429&&\\ \hline
1 & 4                      &    &1&&&\\ 
 & $\frac{7}{2}$  &   1   &14&&&\\ 
 &3                        &14  &91&&&\\ 
 &  $\frac{5}{2}$   & 90&664&&&\\ 
 & 2                      & 350&1000&&&\\ 
 &  $\frac{3}{2}$    & 910&1988&&&\\ 
 & 1                        & 1638&2912&&&\\ 
 & $\frac{1}{2}$       & 2002 & 3068&&&\\ 
 &0                   &1430 &2002 &&&\\ \hline
 0 & 4                 &    1& & & &\\ 
 & $\frac{7}{2}$  &    16& & & &\\ 
 &3                    &   119 & & & &\\ 
 &  $\frac{5}{2}$    &544& & & &\\ 
 & 2                   &    1700& & & &\\ 
 &  $\frac{3}{2}$ &    3808& & & &\\ 
 & 1                   &6188 & & & &\\ 
 & $\frac{1}{2}$  &   7072 & & & &\\ 
 &0                 &   4862 & & & &\\ \hline
\end{tabular}
\end{center}
\normalsize
\begin{center}Table 3: Massive supersymmetry
representations of $N=8$\end{center}
\end{table}
\normalsize}
\begin{table}
\begin{center}
\begin{tabular}{|r |r| r r r|}
\hline $q$&$J$&$L=0$&$L=\frac{1}{2}$& $L=1$\\
\hline\hline
2 & 2                 &     &   & 1     \\ 
 & $\frac{3}{2}$  &    & 1   & 4    \\ 
 &     1              & 1    & 4  &  6  \\ 
 &  $\frac{1}{2}$ & 4   & 6 &  4 \\ 
 &0                    & 5 & 4&   1 \\ \hline
1 & 2                 &     &1   &    \\ 
 & $\frac{3}{2}$  & 1   &6    & \\ 
 &     1              &  6   & 15  & \\ 
 &  $\frac{1}{2}$ &  14  & 20 &    \\ 
 &0                    &14 & 14&\\ \hline
0 & 2                 &  1   &   &      \\ 
 & $\frac{3}{2}$  &  8  &    &      \\ 
 &     1              & 27    &   &    \\ 
 &  $\frac{1}{2}$ & 48   &  &    \\ 
 &0                    &  42& & \\ \hline
\end{tabular}
\end{center}
\normalsize
\begin{center}Table 4: Massive supersymmetry representations
 for $N=4$\end{center}
\end{table}

\begin{table}

\begin{center}
\begin{tabular}{|r |r| r r|}
\hline $q$&$J$&$L=0$&$L=\frac{1}{2}$\\
\hline\hline
1 &     1              &     & 1   \\ 
  &  $\frac{1}{2}$ & 1   & 2 \\ 
 &0                    & 2 & 1\\ \hline
0 &      1              &  1   &    \\ 
 &  $\frac{1}{2}$ &  4  &      \\ 
 &0                    &5 & \\ \hline
\end{tabular}
\end{center}
\normalsize
\begin{center}Table 5: Massive supersymmetry representations
 for $N=2$\end{center}
\end{table}
\normalsize

Let us begin by considering black hole solutions of the $N=8$
supergravity theory with $Z_1 \geq Z_2\geq Z_3 \geq Z_4$. Here we have
the five categories:
\bea
M_{central}&=&Z_1=Z_2=Z_3=Z_4~~~~~~~~~~~~~~~~~~~~~~~~~~q=4 \nn \\
&=&Z_1=Z_2=Z_3>Z_4~~~~~~~~~~~~~~~~~~~~~~~~~~q=3  \nn \\
&=&Z_1=Z_2>Z_3\geq Z_4~~~~~~~~~~~~~~~~~~~~~~~~~~q=2    \\
&=&Z_1=Z_2\geq Z_3\geq Z_4~~~~~~~~~~~~~~~~~~~~~~~~~~q=1 \nn \\
&>&Z_1\geq Z_2 \geq Z_3 \geq Z_4~~~~~~~~~~~~~~~~~~~~~~~~~~q=0  \nn 
\eea
In particular, the $a=\sqrt{3}$ black holes preserve $q=4$
supersymmetries and must belong to short supermultiplets which we
assume to be the maximum spin $J=2$, superspin $L=0$ multiplet
$(1,8,27,48,42)$ appearing in Table 3.  The bound state
interpretation requires that the $a=1$ black holes preserving $q=2$
supersymmetries appear in the product of two $a=\sqrt{3}$
representations.  Ignoring the internal quantum numbers, this product
will decompose into multiplets with $4 \geq J_{max} \geq 2$ as follows

\be
\begin{array}{ll}
1\times\left[ 4\right]\oplus 8\times\left[ \frac{7}{2}\right]
\oplus27\times\left[3\right]\oplus48\times\left[\frac{5}{2}\right]
       \oplus42\times\left[2\right]&~~~q=4\\
1\times \left[4\right] \oplus 6\times \left[\frac{7}{2}\right]
                \oplus14\times\left[3\right]\oplus14\times
\left[\frac{5}{2}\right]&~~~q=3 \\
1\times \left[4\right]\oplus 4\times \left[\frac{7}{2}\right]
                              \oplus 5\times \left[3\right]&~~~q=2 \\
1\times \left[4\right]\oplus 2\times \left[\frac{7}{2}\right]&~~~q=1\\ 
1\times \left[4\right]&~~~q=0
\end{array}
\ee
  
Which of the above possibilities is actually realized, however, will
depend on the charge assignments of the two constituents. Suppose each
is singly charged under just one of the Kaluza-Klein $U(1)$s, then the
bound state will again belong to a $q=4$ multiplet if the $U(1)$s are
the same. On the other hand, if one
carries a Kaluza-Klein charge and the other a winding
charge in the same dimension, 
then we get $q=2$. Since these are precisely the
quantum numbers of the $a=1$ black hole, which has twice the mass of the
$a=\sqrt{3}$ black hole, this is entirely consistent with our hypothesis.
Note, however, that although one might also expect to obtain $q=3$ multiplets,
a closer look at the internal quantum numbers shows that these do not in fact
arise, since the $M_{central}=Z_1=Z_2=Z_3>Z_4$ configuration can never
be achieved with $2$-particle states. 

The same arguments go through for an embedding of the black holes into
$N=4$ or $N=2$ supergravity with an appropriate number of matter
multiplets. For the $N=4$ case the compositions for the product of two
short multiplets are given by \footnote{ Since the same group theory applies,
this suggests that higher superspin multiplets also appear in the spectrum
of global $N=4$ Yang-Mills theories, where traditionally attention is
focussed only on maximum spin $1$.}
 \be \begin{array}{ll} 1\times
\left[2\right]\oplus 4\times \left[\frac{3}{2}\right]
       \oplus5\times\left[1\right]& ~~~q=2\\
1\times \left[2\right] \oplus 2\times \left[\frac{3}{2}\right]
            &~~~q=1 \\
1\times\left[ 2\right]&~~~q=0
\end{array}
\ee
For example, the appearance of the $q=0$ multiplet is consistent with
the interpretation of the non-supersymmetric $a=1$ black hole as a bound state 
of a supersymmetric $a=\sqrt{3}$ positively charged Kaluza=Klein black hole and 
a supersymmetric $a=\sqrt{3}$ negatively charged winding black hole.  Similarly,
for $N=2$ we find 
\be \begin{array}{ll} 1\times\left[ 1\right]\oplus
2\times\left[ \frac{1}{2}\right]
       & ~~~q=1\\
1\times\left[ 1\right]&~~~q=0 
\end{array}
\ee
Therefore, we have shown that the supermultiplet structures are
consistent with our hypothesis in the case of $2$-particle bound states.
It is straightforward to show by taking further tensor products that
things go through in a similar way for the $3$- and $4$-particle states.

It should be stressed that we are not claiming that all bound states can
be identified with oscillator states. In the $N=4$ table, for example, $L=1/2,
J_{max}=3/2$ states appear in the tensor product, whereas there are no $L=1/2$
supermultiplets in the oscillator spectrum of the $N=4$ heterotic string because 
the left-moving sector is purely bosonic.  Similarly, the $L=1,J_{max}=2$
states appearing in the tensor product will be black holes whose angular
momentum is fermionic in origin \cite{Aichelburg}, whereas the $L=1$ oscillator
states have a bosonic Kerr-type angular momentum and cannot therefore be
black holes.

So far we have put ourselves at the special point in moduli space $S=T=U=i$. At
generic points, the mass formula becomes \cite{Ceresole,Rahmfeld3}
\be
M_{central}=\frac{1}{{\rm Im}~S~ {\rm Im}~T~{\rm Im}~U}  
{\rm
max}|(\alpha_1+U\alpha_2+T\alpha_4-TU\alpha_3)\pm(S\beta_1-SU\beta_2-ST\beta_4
-STU\beta_3)| \ee
where $\alpha$ and $\beta$ are the quantized electric and magnetic charge
vectors. Thus, although the bound state interpretation continues to apply, the
zero binding energy phenomenon does not. The triangle inequality ensures that
at generic points the bound states have non-zero binding energy (except of
course when the charges are not relatively prime). 

\section{Multi-black brane solutions of $M$-theory}
\la{multi}

\vspace{1truecm}

\begin{center}
\begin{tabular}{|l||l|l|l|l|}
\hline & 4-form & 3-form & 2-form & 1-form \\
\hline\hline
D=11&4&&&\\ \hline
D=10 &$4$& $4$ &$4$ & \\ \hline
D=9 &$4$&$ 4$ &$4,2$ &$4$ \\ \hline
D=8 &$4$&$ 4$ &$4,2$ &$4,2$ \\ \hline
D=7 & &$4 $ &$4,2$ &$ 4,2$\\ \hline
D=6 && $4,2$ & $4,2$&$4,2,\frac{4}{3},1$ \\ \hline
D=5 && &$4,2,\frac{4}{3}$ &$4,2,\frac{4}{3},1$ \\ \hline
D=4 &&  &$4,2,\frac{4}{3},1$ &$4,2,\frac{4}{3},1,\frac{4}{5},
\frac{2}{3},\frac{4}{7}$  \\ 
\hline
\end{tabular}
\la{brane}
\end{center}
\begin{center}Table 6: $\Delta$ values for 
supersymmetric $p$-branes\end{center}

\vspace{1truecm}
Let us now turn to the black branes of $M$-theory
\cite{Dufflupope2,Klebanov1}. 
Starting from eleven dimension, toroidal compactification gives rise to a
variety of $(d +1)$-form field strengths and hence fundamental
$(d -1)$-brane and solitonic $(\tilde d-1)$-brane solutions in the lower
dimension $D=d+\tilde d+2$.  The compactified eleven-dimensional
supergravity theory admits a consistent truncation to the following set
of fields: the metric tensor $g_{MN}$, a set of $n$ scalars
$\vec{\phi}=(\phi_1,...\phi_n)$, and $n$ field strengths $F_{\alpha}$ of
rank $(\tilde d +1)$.  The $D$-dimensional action describing the
$p$-branes under consideration  is then given by \cite{Popelu2}    
\be
\lag=\sqrt{-g}\left[R - \frac{1}{2} (\partial \vec{\phi})^2 -\frac{1}
{2\times (d+1)!}
  \sum_{\a=1}^{n} e^{\vec {a}_\a \vec{\phi}} F_{d+1}^{\a \ 2}\right]\ , 
\label{actionp}
\ee
where $n$ is the number of participating field strengths. If all active
charges are equal, this can be
further truncated to the Lagrangian (\ref{actiona}) involving a single scalar
and single field strength where $a$, $\phi$ and $F$ are given by 

\[ a^{-2}=\sum_{\alpha\beta}(M^{-1})_{\alpha\beta}\]

\[\phi=a\sum_{\alpha\beta}(M^{-1})_{\alpha\beta}\vec {a}_\a \vec{\phi}\]

\be (F_{\alpha})^2=a^2\sum_{\alpha\beta}(M^{-1})_{\alpha\beta}F^2 \ ,
\ee
where the matrix $ M_{\a \b}$ is given by
\be  M_{\a \b}=\vec{a}_{\a} \vec{a}_{\b}. 
\la{matrix}
\ee
The parameter $a$ can conveniently be expressed as
\be
a^2=\Delta-\frac{2d\tilde d}{d+\tilde d}
\ee
As discussed in \cite{Popelu2},
supersymmetric $p$-branes solutions can arise only when the value of $\Delta$ is
given by $\Delta=4/n$.  This occurs when   \be
 M_{\a \b}=4\delta_{\a \b} -2\ratio.
\ee

Originally, attention was focussed on the
$\Delta=4$ solutions \cite{Strominger1,Dufflu,Khuristring} but various new
supersymmetric solitons with $\Delta\neq 4$ have recently been studied
\cite{Popestainless,Popelu1,Popelu2}. In this section we shall
generalize the treatment of extremal black holes \cite{Rahmfeld4} and
confirm on the level of classical solutions that these new $p$-branes
can also be regarded as bound states with zero binding energy of
fundamental $\Delta=4$ branes.  We find new $1\leq n \leq m$-centered
$p$-brane solutions which reproduce the $\Delta=4/n$  solutions of
\cite{Popelu2} as we allow $n$ of the centers to coincide and take the
remaining $(m-n)$ out to infinity. Table 6 summarizes in which
dimensions the various solitons in maximal supergravities arise. Below
eight (six)  dimensions the four(three)-form field strengths are
dualized. 

For the purposes of exhibiting the multi-centered solutions, we shall work
with (\ref{actionp}).  The equations of motion  are
\bea
  \frac{1}{\sqrt{-g}}\partial_{M}\left(\sqrt{-g}  e^{\vaa \vec\phi} 
F_{d+1}^{\a \  MN}\right)&=&0 
                       \la{eom1}\\
 \frac{1}{\sqrt{-g}}\partial_{M}\left(\sqrt{-g} \partial^{M} \phi_i\right)
&=&
  \sum_{n}\frac{a_{i\a}}{2\times (d+1)!}F_{d+1}^{\a \ 2} 
\la{eom2}\\
  R_{MN}&=&\frac{1}{2}\partial_{M}\vp\partial_{N}\vp + \nn \\
& & \qquad\quad 
+\frac{1}{2 d !}\sum_{n}e^{\vaa \vp}\biggl(F^{\a \
2}_{MN}-\frac{d}{n(d+\dt)}
 F^{\a \ 2}
  g_{MN}\biggr) \la{eom3}
\eea
The one-center solutions of these equations have been intensively studied. 
Here we find the {\it multi-center} solution which reads for the solitonic
case:
\bea
f_{\a}&=&(1+\frac{\l_\a}{d|\vec{y}-\vec{y}_{0,\a}|^{d}}) \la{factor}\\
ds^2&=&\prod_{\a=1}^{n}f_{\a}^{-\frac{d}{d+\dt}}dx^{\mu}dx_{\mu} +
\prod_{\a=1}^{n}f_{\a}^{\frac{\dt}{d+\dt}}dy^m dy_m \\
e^{-\vaa \vp}&=& f_{\a}^2 \prod_{\b=1}^{n}f_\b^{-\frac{d\dt}{d+\dt}} \\
F^\a_{m_1...m_{d+1}}&=&\l_\a\e_{m_1 ...m_{d+1}p} \frac{y^p}{|\vec{y}-
\vec{y}_{0,\a}|^{d+2}} 
\eea
where $\mu$ refers to the $\dt$ world-volume coordinates of the solitonic
$(\dt-1)$-brane and $m$ to the
$D-\dt=d+2$ transverse coordinates. In the special case $d=0$ we 
choose 
\be
f_{\a}=(1+\l_\a\ln{\frac{|\vec{y}-
\vec{y}_{0,\a}|}
{r_{0,\a}}})
\ee
instead of (\ref{factor}). The solutions  for elementary 
multi-$p$-branes are easily obtained by generalizing the
single-centered solutions of \cite{Popelu2} along the lines above. 
Since (\ref{eom1}) and (\ref{eom2}) are
essentially linear in the contributions of each field strength they
are obviously satisfied for the multi-center solutions. The only
slightly non-trivial check comes from the Einstein equations, since
here the scalar fields and $R_{mn}$ involve non-linearities in the
individual soliton contributions. If we plug the ansatz into the
field equations we find the following condition for vanishing
non-linearities: 
\bea \frac{d\dt^2}{2(d+\dt)}+\dt
M^{-1}_d\left[(n-2)(\ratio)^2-2 (2-\ratio) \ratio\right] + \ \ \ \ \ \
\ \ & & \nn \\
\hspace*{-0.5truecm}+\dt M^{-1}_n\biggl[(2-\ratio)^2
+(\ratio)^2-2(n-2) (2-\ratio) \ratio+ \ \ & & \nn \\ +2(n-2)
(\ratio)^2+(n-2)(n-3)(\ratio)^2\biggr]&=&0, 
\eea 
where $M^{-1}_{d(n)}$ denote the (off)-diagonal elements of the matrix
$M_{\a \b}$ defined in (\ref{matrix}). Indeed, for all the new
supersymmetric $p$-branes the above condition holds, allowing us to
generalize the single-center solutions to multi-center ones. However,
there is one subtlety, for certain values of $d,n,D$, $M$ is singular
and cannot be inverted.  The problem arises for ($D=4, n=4, d=1$) and
($D=5, n=3, d=1$). The first case was considered in \cite{Rahmfeld4}
and shown to be a valid solution. The second one can also be shown to 
work by an independent calculation.  If so desired, one may now consider the
special case of coincident centers and equal charges to obtain the
$\Delta=4/n$ solutions, thus confirming that these admit the interpretation
of bound states with zero binding energy of $n$ fundamental $\Delta=4$
branes. 

Most of our results are not based on the fact that we consider maximal
supergravities. So we can ask ourselves what kind of bound states
survive in the heterotic theory. It appears that all $1$-form and $2$-form
types do. In four dimensions we also have the four classic black hole
types and also strings with up to seven participating field
strengths. In the heterotic theory the number seven finds a very
natural explanation in the presence of one $S$-field 3 $T$-fields and
3 $U$-fields of section (2) \cite{Rahmfeld3,Rahmfeld2}. The multi-string
solution  found in \cite{Rahmfeld2} with non-vanishing $S$ and $T$
fields is given by
\bea
ds^2&=&-dt^2+dz^2+S_2T^{(1)}_2T^{(2)}_2T^{(3)}_2(dx^2+dy^2) \nn \\
S&=&S_1+iS_2=\sum_{i=1}^{n}s_i\ln{\frac{x+iy-w_i}{r_{i}}}\\
T^{(a)}&=&T^{(a)}_1+iT^{(a)}_2=
\sum_{i=1}^{n}t^{(a)}_i\ln{\frac{x+iy-w^{(a)}_i}{r^{(a)}_{i}}}
\eea
where $w_i=x_i+iy_i$ and $r_{i}$ denote the positions and sizes of
the sources. The solution
preserves $1/2$, $1/4$ and $1/8$ of the supersymmetries
for one, two and three $T$-fields but without an $S$ charge. In the
presence of the $S$ field the supersymmetries get halved with the
exception of the configuration with 3 $T$ and one $S$ field which 
breaks either breaks either seven eighth or all supersymmetries, depending
on the chirality choice.  It is straightforward to generalize these
results to include the $U$ fields.

The presence of only one 3-form in all dimensions (above five)
forbids the $D=6,d=2,n=2$ solution. Nevertheless, we have a very
interesting solution in $D=6,d=2$, which also can be viewed as a bound
state: the dyonic string of \cite{Rahmfeld2,Dufflupope1}:
\bea
\Phi&=&\Phi_E+\Phi_M, \nn \\
ds^2&=&e^{\Phi_E}(-dt^2+dz^2)+e^{\Phi_M}dx^idx_i \label{dyonstring}\\
e^{-\Phi_E}&=& 1+\frac{Q}{y^2} \qquad 
e^{-\Phi_M}= 1+\frac{P}{y^2} \nn \\
H_3&=& 2Q\e_3+2Pe^{\Phi}*\e_3 \nn
\eea
with $y^2=x^ix_i, \ \ i=1,2,3,4$ and (in general)
\be
e^{\Phi_E}\partial^2 e^{-\Phi_E}=e^{\Phi_M}\partial^2 e^{-\Phi_M}=0
\ee
Since the  electric as well as the
magnetic part are determined by two independent harmonic functions,
we can easily generalize (\ref{dyonstring}) to a multi-string
solution. The same holds for the ten dimensional dyonic string
also found in \cite{Rahmfeld2}:
\bea
ds^2&=&e^{\Phi_{E_1}+\Phi_{E_2}}
(-dt^2+dz^2)+e^{\Phi_{M_1}}\delta^{ij}dy_idy_j
+e^{\Phi_{M_2}}\delta^{ab}dy_ady_b \nn \\
e^{\Phi_{E_\a}} \partial^2_1 e^{-\Phi_{E_\a}}&=&0 \qquad
e^{\Phi_{M_\a}} \partial^2_2 e^{-\Phi_{M_\a}}=0 \la{tendyon}
\eea
with $i,j=2,3,4,5; \ \ a,b=6,7,8,9$ and $\a=1,2$. $\partial^2_{1,2}$
denote the d'Alemberts in the $(2,3,4,5)$ and $(6,7,8,9)$ subspaces
respectively. The antisymmetric
tensor field is given as 
\be
B_{01}=e^{\Phi_{E_1+E_2}}, \ \ H_{ijk}=\e_{ijkl}\partial^l\Phi_{M_1},
\ \ H_{abc}=\e_{abcd}\partial^d\Phi_{M_2}.
\ee
(\ref{tendyon}) is written in ten dimensional string coordinates.

Another interesting aspects of the multi-$p$-brane solutions is that
the charge parameters of the individual constituents are
independent. In particular, if we have only two $p$-branes, we can
choose their charges to be of equal magnitude but different sign, which
leads to massless solutions
\cite{Cveticyoum2,Behrndt,Kalloshcancel,Chancvetic,Popelu1,Ortin,Dufflupope1}. 
The interpretation is that two supersymmetric $p$-branes, one with positive and
one with negative mass, can combine to form a supersymmetric massless
$p$-brane. Certainly, an isolated brane with negative mass does not make sense
quantum-mechanically but there may be some quantum confinement mechanism
that allows it to exist only as a bound state.

\section{Entropy and temperature}
\la{entropy}

In this section, we ask whether the entropy and temperature of these
black $p$-branes \cite{Dufflupope2,Klebanov1,Mukherji2} is consistent with the
bound state interpretation.  The mass per unit volume and the charges of the
multi black $p$-branes may be written in terms of parameters $k$ and
$\mu_{\alpha}$ as \cite{Dufflupope2}
\[
M_{black}=k(\tilde d \sum_{\a=1}^{n}\sinh^2\mu_{\alpha} +\tilde d +1)   
\]
\be
\lambda_{\alpha}=\frac{1}{2}\tilde d k \sinh 2\mu_{\alpha}
\ee
The Hawking temperature $T$ and entropy $S$ of these multi black
$p$-branes in the case where all centers are coincident are given by
\cite{Dufflupope2} 
\[
T=\frac{\tilde d}{4\pi y_+} \prod_{\a=1}^{n}(\cosh\mu_{\alpha})^{-1}
\]
\be
S=\frac{1}{4}{y_+}^{\tilde d +1} \omega_{\tilde d +1}
\prod_{\a=1}^{n}(\cosh\mu_{\alpha})
\ee
where the event horizon is located at $y={y_+}=k^{1/\tilde d}$ and where
$\omega_{\tilde d +1}$ is the volume of the unit $(\tilde d +1)$-sphere.

The form of the total entropy as a product of the individual entropies is
puzzling from the  bound state interpretation. It remains a puzzle when we take
the extremal limit. To illustrate this, let us consider the special case where
each of the $n$ field strengths is equal. Then we have
\cite{Dufflupope2}
\[
T=\frac{\tilde d}{4\pi y_+}(\cosh\mu)^{-n}
\]
\be
S=\frac{1}{4}{y_+}^{\tilde d +1} \omega_{\tilde d +1}(\cosh\mu)^n
\ee
The extremal limit corresponds to $k\rightarrow 0$, $\mu\rightarrow
\infty$, holding $\lambda=\tilde d \sqrt{n/2}ke^{\mu}$ constant. Thus
the entropy vanishes unless the constant $a$ is zero and $d=1$. This
happens only for $(D=4,n=4,d=1)$, which is just the Reissner Nordstrom
black hole, and for the five-dimensional black hole $(D=5,n=3,d=1)$. 
Remarkably, these were precisely the two cases where the matrix $M$
of (\ref{matrix}) was singular.  At first sight this result seems strange: In
$D=4$, for example, it seems very unnatural to combine three black holes and
still the entropy is zero; but adding a fourth one suddenly forces it to be
finite. To see in more detail how this comes about, let us 
invoke \cite{Ferrarakallosh}, where a new recipe for the
calculation of the horizon area was given: the scalar fields on the horizon
(in our case $\eta$, $\sigma$ and $\rho$) are determined by the requirement
that the central charge becomes extremal. The entropy is then given by the
value of the central charge evaluated with the scalar fields at the
horizon.  For the standard black holes with charges $Q_1,Q_3,P_2$ and $P_4$ the
scalar fields on the horizon are fixed to \cite{Ferrarakallosh} \be 
e^{-2\eta}\rightarrow|\frac{P_2 P_4}{Q_1Q_3}| \ , \ \ \ \
 e^{-2\s}\rightarrow|\frac{P_2Q_3}{Q_1 P_4}| \ ,\ \ \ \
e^{-2\r}\rightarrow|\frac{Q_3 P_4}{Q_1 P_2}|.
\ee
For each ``incomplete'' black hole, i.e. a state with not all four
charges non-zero, at least one of the scalars blow up, either to plus
or minus infinity. This ties in very nicely with our bound state
hypothesis. For an $a=1/{\srt}$ with (for example)
$Q_1=Q_3=P_2=1$ all three scalars diverge: 
\be
\eta\rightarrow\infty \ , \ \ \ \ \s\rightarrow-\infty \ , \ \ \ \
\r\rightarrow \infty.  \ee The elementary magnetic black hole with
$P_4=1$ on the other hand has diverging scalars with the opposite
sign: \be \eta\rightarrow-\infty \ , \ \ \ \ \s\rightarrow\infty \ , \
\ \ \ \r\rightarrow-\infty!  
\ee 
Since we know from \cite{Rahmfeld4,Cvetictseytlin} that there is a
multi-black hole solution and the scalars are additive we can safely
conclude that the $a=1/\sqrt{3}$ and $a=\sqrt{3}$ state 
conspire to force the scalars and therefore the
entropy to be finite.

All this suggests that it is only the $a=0$ non-dilatonic $p$-brane
entropies that can be trusted since when the dilaton is non-trivial
there will be strong coupling effects that we do not yet know how to
handle.  However, it was shown in \cite{Mukherji2} that if quantum corrections
smooth  out singularities, even black holes with $a \neq 0$ may have
non-vanishing entropy.   

\section{Conclusions}

We have reexamined the suggestion in \cite{Rahmfeld1} that, in the
extremal limit, the non-rotating black hole solutions of string theory may
be identified with both elementary and solitonic BPS string states, and
have confirmed that the interpretation of certain multiply charged black
holes as bound states at threshold of singly charged black
holes is consistent with the masses, charges, spins and supermultiplet
structure of the string states. We also confirm that the bosonic Kerr-type
angular momentum, arising from the left-moving sector of the heterotic
string, corresponds to the superspin $L$ of the oscillator states and hence that
only $L=0$ BPS oscillator states can be black holes. One is tempted to conclude
that the $L>0$ oscillator states must therefore be described by naked
singularities, but there also exist solutions that are asymptotically identical
to such solutions but near the core have a much milder singularity and whose
angular momentum is naturally Regge-bounded \cite{Harveyblack}.  Moreover, this
bound state interpretation generalizes to super $p$-branes in $D$ dimensions. In
doing so, of course we have to put ourselves at special points in moduli space.
In section (\ref{bound}), for example, we set the asymptotic value of ${\cal M}$
to $I$ and the asymptotic value of the dilaton to zero. For generic points
in moduli space, the bound state interpretation would continue to apply but we
would no longer have the zero binding energy phenomenon (except of course if
the charges are not relatively prime). We might add that these results are also
consistent with the recent recognition that some $p$-branes carrying
Ramond-Ramond charges admit an interpretation as Dirichlet-branes, or
$D$-branes, and are therefore amenable to the calculational power of conformal
field theory \cite{Polchinski}. Bound states of $p$-branes have been discussed
from the somewhat different perspective of Dirichlet branes in
\cite{Schwarz3,Witten2,Li,Sen3,Vafa,Tseytlin}.  Apart from their intrinsic
importance, therefore, these black holes and black $p$-branes have recently come
to the fore as away of providing a microscopic explanation of the Hawking
entropy and temperature formulae which have long been something of an enigma.
See \cite{Horowitz} for a recent review. For reasons discussed in section
(\ref{entropy}), however, only the $a=0$ branes are amenable to these
calculations, given our current technology.

An interesting special case is provided by the $a=\sqrt{3},1,1/\sqrt{3},0$
black holes which admit the interpretation as $1,2,3,4$-particle bound
states at threshold \cite{Rahmfeld1,Rahmfeld3}. One feature which appeared
mysterious to us at the time was: Why the unit charge solutions singled
out {\it four} values of $a$ and hence why only $1,2,3,4$- and
not $5,6,...$-particle bound states? An explanation of this has recently been
given  \cite{Townsendpapa,Klebanov2,Behrndt2,Gauntlett,Larsen} in terms of
intersecting membranes \cite{Duffstelle} and fivebranes \cite{Guven} in 
$D=11$. This opens up a new direction for the study of black hole and black
$p$-brane bound states.      

Finally, another crucial consistency check on the black hole, bound state,
string state picture is supplied by comparing gyromagnetic and
gyroelectric ratios.  This turns out to be quite subtle \cite{Russo} and will
treated in a separate publication \cite{Duffliurahmfeld}.
   
\section{Acknowledgements}

We have enjoyed useful conversations with Jim Liu and Massimo Porrati
on string state gyromagnetic ratios and with Sudipta Mukherji, Hong Lu and
Chris Pope on $p$-brane entropy.

\newpage

\bibliographystyle{preprint}

\begin{thebibliography}{10}

\bibitem{Rahmfeld1}
M.~J. Duff and J.~Rahmfeld,
\newblock {\sl Massive string states as extreme black holes},
\newblock Phys. Lett. {\bf B 345} (1995) 441.

\bibitem{Sen2}
A.~Sen,
\newblock {\sl Extremal black holes and elementary string states},
\newblock Mod. Phys. Lett. {\bf A 10} (1995) 2081.

\bibitem{Peet}
A. Peet,
\newblock {\sl Entropy and supersymmetry of D dimensional extremal black
holes versus string states},
\newblock hep-th/9506200.

\bibitem{Khurimyers}
R. R. Khuri and  R. C. Myers,
\newblock {\sl Dynamics of extreme black holes and massive string states}, 
\newblock hep-th/9508045.

\bibitem{Callan}
C. G. Callan, J. M. Maldacena and A. W. Peet,
\newblock {\sl  Extremal black holes as fundamental strings}, 
\newblock hep-th/9510134.

\bibitem{Shiraishi}
K. Shiraishi,
\newblock {\sl Extreme dilatonic black holes on a torus}, 
\newblock hep-th/9510134.

\bibitem{Gibbons}
G.~W. Gibbons,
\newblock {\sl Antigravitating black hole solitons with scalar hair in $N=4$
supergravity}, 
\newblock Nucl. Phys. {\bf B 207} (1982) 337.

\bibitem{Khurinew}
M.~J. Duff, R.~R. Khuri, R.~Minasian and J.~Rahmfeld,
\newblock {\sl New black hole, string and membrane solutions of the four
  dimensional heterotic string},
\newblock Nucl. Phys. {\bf B 418} (1994) 195.

\bibitem{Schwarz1}
J.~H. Schwarz and A.~Sen,
\newblock {\sl Duality symmetries of 4-{$D$} heterotic strings},
\newblock Phys. Lett. {\bf B 312} (1993) 105.

\bibitem{Duffpope}
 M. J. Duff and C. N. Pope,
\newblock {\sl Consistent truncations in Kaluza-Klein theory},
\newblock Nucl. Phys. {\bf B255} (1985) 355.

\bibitem{Pope}
C. N. Pope,
\newblock {\sl The embedding of the Einstein Yang-Mills equations in
$D=11$ supergravity},
\newblock Class. Quantum Grav. {\bf 2} (1985) L77.



\bibitem{Horowitz2}
G.~W. Gibbons, G.~T. Horowitz and P.~K. Townsend,
\newblock {\sl Higher dimensional resolution of dilatonic black hole
  singularities},
\newblock Class. Quantum Grav. {\bf 12} (1995) 297.

\bibitem{Rahmfeld3}
M.~J. Duff, J.~T. Liu and J.~Rahmfeld,
\newblock {\sl Four dimensional string/string/string triality},
\newblock Nucl. Phys. {\bf B 459} (1996) 125.

\bibitem{Cveticyoum}
M.~{Cveti\v c} and D. Youm,
\newblock {\sl Dyonic BPS saturated black holes of heterotic string
theory}, 
\newblock {\tt hep-th/9510098}.


\bibitem{Cvetictseytlin}
M.~{Cveti\v c} and A.~A. Tseytlin,
\newblock {\sl Solitonic strings and bps saturated dyonic black holes},
\newblock {\tt hep-th/9512031}.

\bibitem{Kalloshpeet}
R.~Kallosh, A.~Linde, T.~Ortin, A.~Peet and A.~V. Proeyen,
\newblock {\sl Supersymmetry as a cosmic censor},
\newblock Phys. Rev. {\bf D 46} (1992) 5278.

\bm{Hull}
C.~M. Hull and P.~K. Townsend,
\newblock {\sl Unity of superstring dualities},
\newblock Nucl. Phys. {\bf B 438} (1995) 109.

\bibitem{Popelu1}
H.~Lu and C.~N. Pope,
\newblock {\sl P-brane solitons in maximal supergravities},
\newblock {\tt hep-th/9512012} .




\bibitem{Khuriortin}
R.~R. Khuri and T.~Ortin,
\newblock {\sl Supersymmetric black holes in $N=8$ supergravity},
\newblock {\tt hep-th/9512177}.

\bibitem{Rahmfeld4}
J.~Rahmfeld,
\newblock {\sl Extremal black holes as bound states},
\newblock Phys. Lett. {\bf B 372} (1996) 198.

\bibitem{Kalloshcancel}
R.~Kallosh and A.~Linde,
\newblock {\sl Supersymmetric balance of forces and condensation of bps
  states},
\newblock {\tt hep-th/9511115}.

\bibitem{Holzhey}
C.~F. Holzhey and F.~Wilczek,
\newblock {\sl Black holes as elementary particles},
\newblock Nucl. Phys. {\bf B 380} (1992) 447.

\bibitem{Strominger2}
A.~Strominger,
\newblock {\sl Massless black holes and conifolds in string theory},
\newblock Nucl. Phys. {\bf B 454} (1995) 96.

\bibitem{Behrndt}
K.~Behrndt,
\newblock {\sl String duality and massless string states},
\newblock {\tt hep-th/9510080}.

\bibitem{Cveticyoum2}
M.~{Cveti\v c} and D. Youm,
\newblock {\sl Singular {BPS} saturated states and enhanced symmetries of
four- dimensional $N=4$ supersymmetric string vacua},
\newblock Phys. Lett. {\bf B 359} (1995) 87.

\bibitem{Chancvetic}
K.~L. Chan and M.~{Cveti\v c},
\newblock {\sl Massless {BPS} saturated states on the two torus moduli 
subspace of heterotic string},
\newblock {\tt hep-th/9512188}.

\bibitem{Garfinkle}
D.~Garfinkle, G.~T. Horowitz and A.~Strominger,
\newblock {\sl Charged black holes in string theory},
\newblock Phys. Rev. {\bf D 43} (1991) 3140.

\bibitem{Howe}
M. J. Duff, P. S. Howe, T. Inami and K. S. Stelle,
\newblock {\sl Superstrings in $D=10$ from supermembranes in $D=11$}
\newblock Phys. Lett. {\bf B191} (1987) 70. 

\bibitem{Townsend}
P. K. Townsend,
\newblock  {\sl The eleven dimensional supermembrane revisited},
\newblock Phys. Lett. {\bf B 350} (1995) 184.

\bibitem{Witten}
E. Witten,
\newblock  {\sl String theory dynamics in various dimensions},
\newblock Nucl. Phys. {\bf B 443} (1995) 85.

\bibitem{Duffliuminasian}
M.~J. Duff, J.~T. Liu and R. Minasian,
\newblock {\sl Eleven dimensional origin of string/string duality: a one loop
test},  \newblock Nucl. Phys. {\bf B 459} (1996) 125.

\bibitem{Schwarz2}
J. H. Schwarz,
\newblock  {\sl The power of $M$ Theory},
\newblock Phys. Lett. {\bf B 367} (1996) 97.

\bibitem{Horava}
P. Horava and E. Witten, 
\newblock {\sl Heterotic and type $I$ string dynamics from eleven
dimensions},  
\newblock Nucl. Phys. {\bf B 460} (1996) 506.

\bibitem{Dufflupope2}
M. J. Duff, H. Lu and C. N. Pope,
\newblock {\sl The black branes of $M$-theory},
\newblock {\tt hep-th/9604052}.

\bibitem{Klebanov1}
I. R. Klebanov and A. A. Tseytlin, 
\newblock{\sl Entropy of near-extremal black $p$-branes},
\newblock {\tt hep-th/9604089}.

\bibitem{Mukherji2}
H. Lu, S. Mukherji, C. N. Pope and J. Rahmfeld,
\newblock {\sl Loop corrected entropy of near-extremal dilatonic $p$-branes},
\newblock {\tt hep-th/9604127}

\bibitem{Muto}
T. Muto,
\newblock {\sl On thermodynamics of black $p$-branes},
\newblock {\tt hep-th/9605017}.

\bibitem{Popestainless}
H.~Lu, C.~N. Pope, E.~Sezgin and K.~S. Stelle,
\newblock {\sl Stainless super $p$-branes},
\newblock {\tt hep-th/9508042}.

\bibitem{Strominger1}
G. T. Horowitz and A. Strominger,
\newblock {\sl Black strings and $p$-branes},
\newblock Nucl. Phys. {\bf B 360} (1991) 197.

\bibitem{Dufflu}
M.~J. Duff and J.~X. Lu,
\newblock {\sl Black and super $p$-branes in diverse dimensions},
\newblock Nucl. Phys. {\bf B 416} (1994) 301.

\bibitem{Khuristring}
M.~J. Duff, R.~R. Khuri and J.~X. Lu,
\newblock {\sl String solitons},
\newblock Phys. Rep. {\bf 259} (1995) 213.


\bibitem{Popelu2}
H.~Lu and C.~N. Pope,
\newblock {\sl Multiscalar $p$-brane solitons},
\newblock {\tt hep-th/9512153}.


\bibitem{fifteen}
M. J. Duff,
\newblock {\sl Supermembranes: the first fifteen weeks},
\newblock Class. Quantum. Grav. {\bf 5} (1988) 189.

\bibitem{Sen}
A.~Sen,
\newblock {\sl Strong-weak coupling duality in four-dimensional string
theory}, 
\newblock Mod. Phys. Lett {\bf A 9} (1994) 3707.

\bibitem{Gibbonsperry}
G.~W. Gibbons and M.~J. Perry,
\newblock {\sl Soliton - supermultiplets and Kaluza-Klein theory},
\newblock Nucl. Phys. {\bf B 248} (1984) 629.

\bibitem{Aichelburg}
P.~C. Aichelburg and F.~Embacher,
\newblock {\sl The exact superpartners of $N=2$ supergravity solitons},
\newblock Phys. Rev. {\bf D 34} (1986) 3006.

\bibitem{Ceresole}
A. Ceresole, R. D'Auria, S. Ferrara and A. van Proeyen,
\newblock {\sl Duality transformations in supersymmetric Yang-Mills theories
coupled to supergravity}, 
\newblock {\tt hep-th/9502072}.

\bibitem{Ferrarasavoy}
S.~Ferrara and C.~A. Savoy,
\newblock {\sl Representations of extended supersymmetry on one and two
  particle states},
\newblock in {\it Supergravity 81}, (Eds. Ferrara and Taylor, C. U. P,1982).

\bibitem{Rahmfeld2}
M.~J. Duff, S.~Ferrara, R.~R. Khuri and J.~Rahmfeld,
\newblock {\sl Supersymmetry and dual string solitons},
\newblock Phys. Lett. {\bf B 356} (1995) 479.

\bibitem{Dufflupope1}
M.~J. Duff, H. Lu and C. N. Pope,
\newblock {\sl Heterotic phase transitions and singularities of the gauge
dyonic string}, 
\newblock {\tt hep-th/9603037}.

\bibitem{Ortin}
T.~Ortin,
\newblock {\sl Massless black holes as black diholes and quadruholes},
\newblock {\tt hep-th/9602067}.

\bibitem{Ferrarakallosh}
S.~Ferrara and R.~Kallosh,
\newblock {\sl Supersymmetry and attractors},
\newblock {\tt hep-th/9602136}.

\bibitem{Harveyblack}
A.~Dabholkar, J.~P. Gauntlett, J.~A. Harvey and D.~Waldram,
\newblock {\sl Strings as solitons and black holes as strings},
\newblock {\tt hep-th/9511053}.

\bibitem{Polchinski}
J. Polchinski,
\newblock{\sl Dirichlet branes and Ramond-Ramond charges},
\newblock {\tt hep-th/9510017}.

\bibitem{Schwarz3}
J. H. Schwarz,
\newblock{\sl An $SL(2,Z)$ multiplet of Type $II$ superstrings},
\newblock Phys. Lett. {\bf B360} (1995) 96.

\bibitem{Witten2}
E. Witten,
\newblock{\sl Bound states of strings and $p$-branes},
\newblock Nucl. Phys. {\bf B460} (1996) 335.
   
\bibitem{Li}
M. Li,
\newblock{\sl Boundary sates of $D$-branes and Dy-strings},
\newblock{\tt hep-th/9510161}

\bibitem{Sen3}
A.~Sen,
\newblock {\sl A note on marginally stable bound states in Type $II$
string theory}, 
\newblock {\tt hep-th/9510229}.

\bibitem{Vafa}
C. Vafa,
\newblock{\sl Gas of $D$-branes and Hagedorn density of BPS states},
\newblock {\tt hep-th/96511088}.

\bibitem{Tseytlin}
 A. Tseytlin, 
\newblock{\sl Harmonic superposition of $M$-branes},
\newblock {\tt hep-th/9604035}.

\bibitem{Horowitz}
G. T. Horowitz,
\newblock {\sl The origin of black hole entropy in string theory},
\newblock {\tt gr-qc/9604051}.

\bibitem{Townsendpapa}
 G. Papadopoulos and P. K. Townsend,
\newblock{\sl Intersecting $M$-branes},
\newblock {\tt hep-th/9603087}

\bibitem{Klebanov2}
I. R. Klebanov and A. A. Tseytlin, 
\newblock{\sl Intersecting $M$-branes as four-dimensional black holes},
\newblock {\tt hep-th/9604166}.

\bibitem{Behrndt2}
K. Behrndt, E. Bergshoeff and B. Janssen,
\newblock {\sl Intersecting $D$-branes in ten and six dimensions},
\newblock {\tt hep-th/9604168}.

\bibitem{Gauntlett}
J. Gauntlett, D. Kastor and J. Transchen,
\newblock {\sl Overlapping branes in $M$-theory},
\newblock {\tt hep-th/9604179}.

\bibitem{Larsen}
V. Balasubramanian and F. Larsen,
\newblock {\sl On $D$-branes and black holes in four dimensions}
\newblock {\tt hep-th/9604189}

\bibitem{Duffstelle}
M. J. Duff and K. S. Stelle,
\newblock {\sl Multimembrane solutions of $D=11$ supergravity},
\newblock Phys. Lett. {\bf B 253} (1991) 113.

\bibitem{Guven}
R. Guven,
\newblock {\sl Black $p$-brane solutions of $D=11$ supergravity theory},
\newblock Phys. Lett. {\bf B 276} (1992) 49.

\bibitem{Russo}
J. Russo and L Susskind,
\newblock {\sl Asymptotic level density in heterotic string theory},
\newblock Nucl. Phys. {\bf B 437} (1995) 611.

\bibitem{Duffliurahmfeld}
M. J. Duff, James T. Liu and J. Rahmfeld,
\newblock (in preparation)






\end{thebibliography}


\end{document}